\documentclass{sf2a-conf2010}
\usepackage{graphicx}
\usepackage{hyperref}
\usepackage[]{natbib}
\usepackage{amssymb}
\usepackage{amsmath}

\newcommand{\ba}{\begin{eqnarray}}
\newcommand{\ea}{\end{eqnarray}}
\def\spose#1{\hbox to 0pt{#1\hss}}

\def\lta{\mathrel{\spose{\lower 3pt\hbox{$\mathchar"218$}}
     \raise 2.0pt\hbox{$\mathchar"13C$}}}
\def\gta{\mathrel{\spose{\lower 3pt\hbox{$\mathchar"218$}}
     \raise 2.0pt\hbox{$\mathchar"13E$}}}
\newcommand{\be}{\begin{equation}}
\newcommand{\ee}{\end{equation}}
\newcommand{\bea}{\begin{eqnarray}}
\newcommand{\eea}{\end{eqnarray}}

\def\setR{\mathbb{R}}
\def\setC{\mathbb{C}}

\newcommand{\nS}{n_{_{\mathrm S}}}
\newcommand{\ureh}{\mathrm{reh}}
\newcommand{\GeV}{\mbox{GeV}}

\begin{document}
\TitreGlobal{SF2A 2010}
%
\title{Fundamental physics in observational cosmology}
\author{Patrick Peter}\address{${\cal G}\setR\varepsilon\setC{\cal O}$
  -- Institut d'Astrophysique de Paris, UMR7095 CNRS, Universit\'e
  Pierre \& Marie Curie,\\ 98 bis boulevard Arago, 75014 Paris, France}
\runningtitle{Fundamental physics in observational cosmology}
%
\setcounter{page}{1}
\index{Peter, P.}

\maketitle
\begin{abstract}
%
%
I discuss, through a few examples, how observational cosmology can
provide insights on hypothetical fundamental physics phenomena or
mechanisms, such as Grand Unified Theory, Superstring alternatives to
the inflation paradigm, and inflation itself.
\end{abstract}
\section{Introduction}

Cosmology has definitely entered a phase of precision measurements:
Cosmic Microwave Background (CMB) anisotropies \citep{WMAP7}, Large
Scale Structure (LSS) observations \citep{LSS} and the
magni\-tu\-de-redshift distribution of Supernov\ae{} Ia \citep{SNIa}
to name the most prominent, have radically transformed the
field. Forthcoming experiments (e.g. Planck) and new data such as
Baryonic Acoustic Oscillations \citep{BAO} or those based on the 21 cm
transition \citep{21cm} will further change our view of primordial
cosmology \citep{PU09}. It is no longer enough to try and roughly
understand cosmological observations: time has come to use these data
in a new way instead of merely gather them.

One such way is to test inflationary predictions in greater details to
decide which version is the most satisfactory, and hence to known how
it should be implemented in a high energy physics scheme, in
particular in a string theory framework; I present in
Sec.~\ref{SecInf} a specific example based on the reheating
temperature \citep{JMCR10}. Another way consists in evaluating
directly some consequences of high energy models such as Grand Unified
Theories (GUT); for this, I concentrate (Sec.~\ref{SecGUT}) on the
still possible fraction of topological defects as active seeds for CMB
fluctuations \citep{BPRS,FRSB,Pog}. Finally, taking the special case
of bouncing cosmology (Sec.~\ref{SecBounce}), I argue that challengers
to inflation may still be worth investigating, both at the theoretical
and observational levels \citep{PPNPN}.

\section{Inflation}\label{SecInf}

Inflation \citep{JM08} is nowadays the most widely accepted solution
to the old puzzles of standard cosmology. Its basic predictions
concern the background itself (e.g., the ratio of the total density
$\rho_\mathrm{tot}$ to the critical density $\rho_\mathrm{crit}$,
$\Omega$ that is expected to be unity to an amazing precision, leading
to a vanishing curvature $\Omega=1\Longleftrightarrow\mathcal{K}=0$)
as well as its perturbations that have acted as primordial seeds for
the formation of the now observed LSS. These perturbations imprinted
the CMB in a way that must be correlated with LSS: a consistent model
should explain both sets of data.

Inflation is modeled by a scalar field slowly rolling in a
potential. A few parameters of this potential allow for an almost
complete fit of the whole available set of data; in fact, it is often
argued that only the slow-roll phase is probed by astrophysical
observations; this phase requires the scalar field to be far from its
equilibrium value. In itself, this is already quite an achievement,
and a way to discriminate between various models.

\begin{figure}[t]
 \centering
 \includegraphics [width=0.6\textwidth,clip]{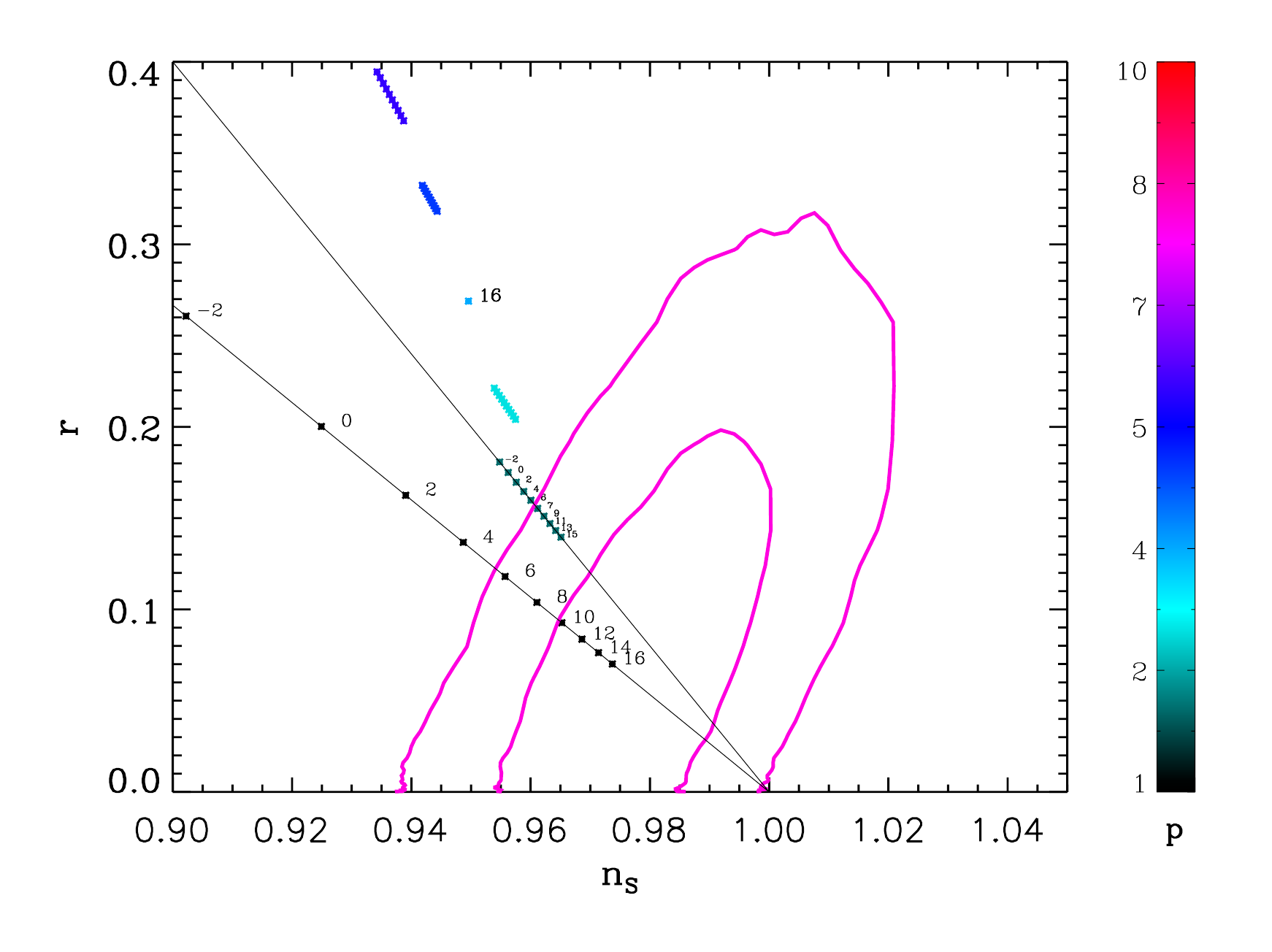}
 \caption{One and two-$\sigma$ WMAP confidence intervals in the
   $(\nS,r)$ plane. Quoted numbers indicate the reheating temperature
   in the form $T_{\ureh}=10^\mathrm{\#} g_*^{-1/4}\GeV)$ for various
   values of the scalar field power $p$ in the potential (lines are
   for $p\gtrsim 1$ and $p=2$). See \cite{JMCR10}, from which this figure is
  taken, for details.}
  \label{FigJMCR}
\end{figure}

The degree of accuracy of the most recent data has now increased to a
such a level that it has become possible to probe the different part
of the potential corresponding to reheating, i.e. the transition
between the end of inflation itself and the radiation dominated era,
at a time for which the scalar field is therefore close to its true
minimum. The reheating phase duration affects the observational range
of scales at which one observes the end of the slow-roll
phase. Measurements of the spectral index $\nS$ and the
tensor-to-scalar ratio $r$ thus constrain this phase characteristics
(equation of state, duration and hence
temperature). Fig.~\ref{FigJMCR}, taken from \cite{JMCR10},
illustrates this point in the case of large field models, for which
the potential takes the form $V\propto \varphi^p$.

\section{GUT and cosmic defects}\label{SecGUT}

Another direct consequence of high energy theories comes straight from
GUT, and those are topological defects \citep{VilShe00} and cosmic
strings in particular, whose formation is almost inevitable in most
cosmologically consistent models \citep{rocher}.

\begin{figure}[ht!]
 \centering
 \includegraphics [width=0.6\textwidth,clip]{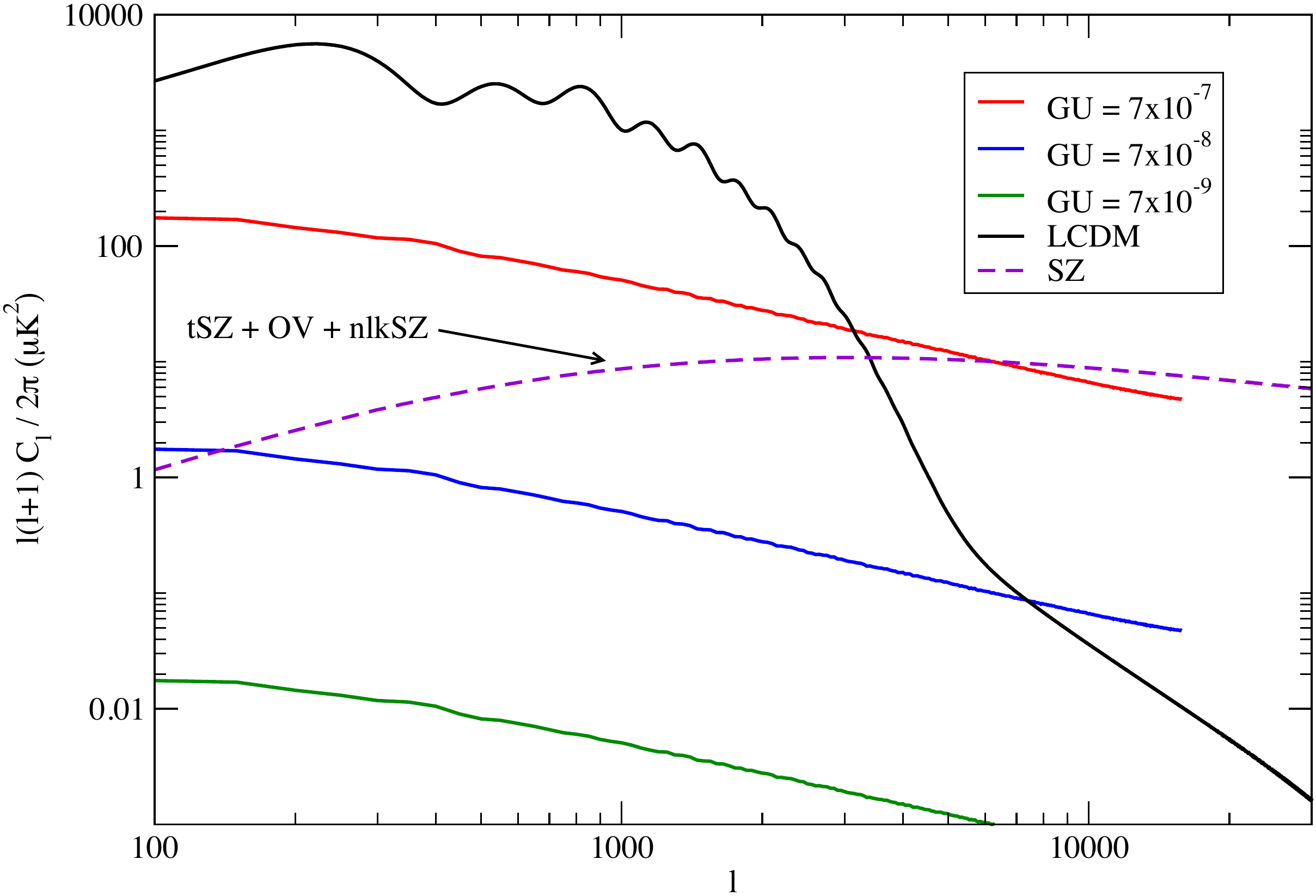}
 \caption{Multipole spectrum for the standard $\Lambda$CDM model --
   best fit to the currently available data \citep{WMAP7} -- and for a
   network of cosmic strings for different values of the energy per
   unit length $U$ in units of the Planck mass squared. The dashed
   curve represents the contribution from the Sunyaev-Zeldovitch
   effects, namely thermal (tSZ), Ostriker-Vichniak (OV) and nonlinear
   kinetic (nlkSZ). From \cite{CR10}}.
  \label{FigInfCS}
\end{figure}

If one considers cosmic string contribution to the CMB as a viable
source of cosmological perturbations, one rapidely finds that it
cannot be made, by itself, to fit the currently available data; this
is due, in particular, to the fact that defects are active seeds, and
hence cannot produce a coherent spectrum. In other words, the
observed acoustic oscillations in the resulting spectrum would be
damped. Fig. \ref{FigInfCS} shows a typical spectrum due to cosmic
strings, as calculated e.g. by \cite{CR10}: those
can merely be part of the final spectrum, although for some values of
the parameters, it is seen that there exists a (small) multipole window
for which they might actually dominate the signal.

As a source of cosmological perturbations, strings would also produce
non gaussianities in a way that is sufficiently different from the
inflation case to be distinguished \citep{CR10}. It can be argued that
forthcoming data have a large potential to exhibit such a string
signal: that would be a direct measurement of the GUT symmetry
breaking energy scale!

\section{Bounces}\label{SecBounce}

As we have seen, the now standard inflationary paradigm can be
modified by inclusion of topological defects. However, the fact that
this mechanism is the best currently available to fit all the data
doesn't mean it is unique. In fact, a reasonnable way to test its
effectiveness would be to find a challenger, and possibly confirm or
rule it out. This is what happened to cosmic strings, so the question
can be asked as to whether there does exist any plausible contender?
It turns out there can be: the bouncing scenario.

\begin{figure}[ht!]
 \centering
 \includegraphics [width=0.45\textwidth,clip]{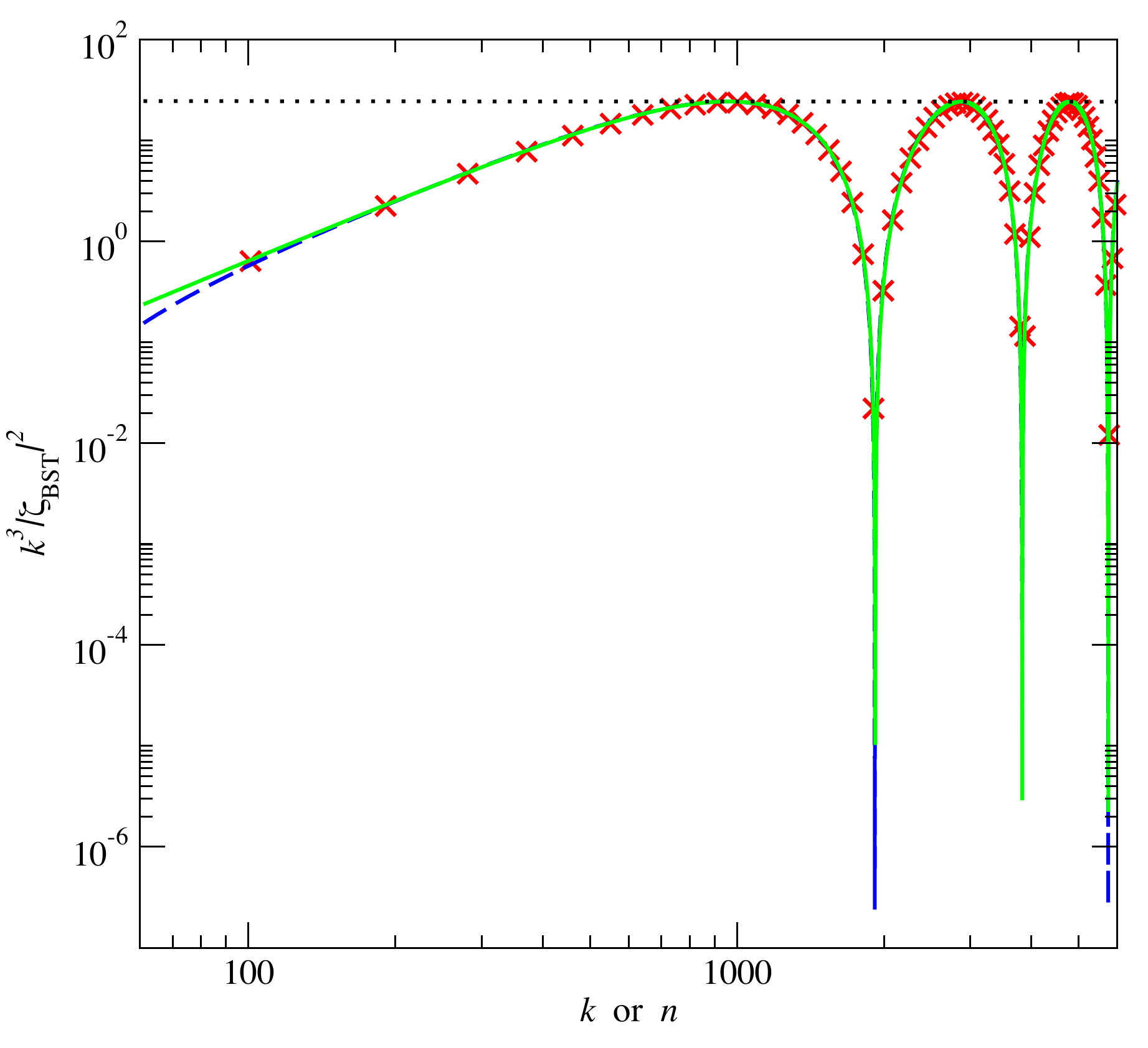}\hskip4mm
 \includegraphics [width=0.45\textwidth,clip]{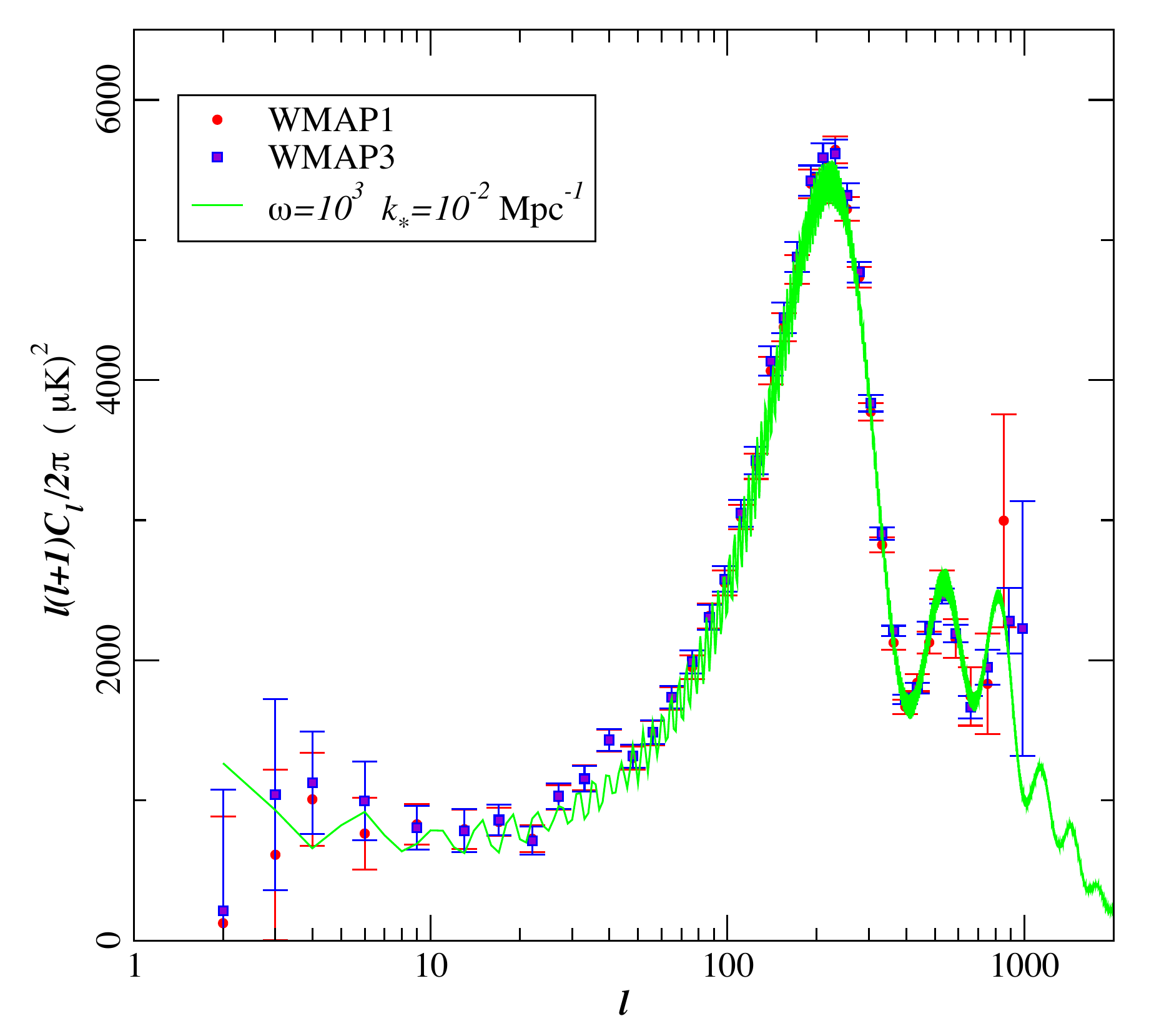}
 \caption{The primordial spectrum $k^3|\zeta_\mathrm{BST}|^2$ (right)
   and subsequent multipoles $C_\ell$ (left) for a  typical bouncing
   model. From \cite{Lilley08}}.
  \label{FigFLP}
\end{figure}

Instead of using a phase of accelerated ($\ddot{a}>0$) expansion
($\dot{a}>0$) as inflation does, having a bounce supposes a phase of
decelerated ($\ddot{a}<0$) contraction ($\dot{a}<0$). As a result,
just before the bounce itself, the total density can be as close to
critical as one wishes. With a contraction dynamics dominated by a
regular, matter or radiation, fluid, the horizon, being an integral
quantity over time \citep{MP04}, can be made to diverge, hence solving
this puzzle as well. The remaining usual issues such as homogeneity
can be addressed under reasonnable additional assumptions
\citep{PPNPN}.

Crucial to any cosmologically relevant model is the existence of
primordial perturbations which will seed the LSS formation; these can
be rather problematic in a bounce model \citep{PPNPN02}. It is
often said that this category of models faces the difficulty of being
very much model dependent which, it is argued, is not the case of
inflation. Not only is this last assertion erroneous, each
inflationary model leading to different observational
prediction\footnote{The fact that it is possible to identify a generic
  ``slow-roll'' category of inflationary models does not mean these
  models produce undistinguishable predictions \citep{MRT10}.}, but it
is also misleading since most bouncing models do also share generic
prediction features.

The most important difficulty faced by bouncing models is that
classical General Relativity (GR) is very reluctant, under general
energy conditions, to let the Hubble rate vanish! Options however can
be considered. In the framework of GR, one needs at least a positive
spatial curvature and a scalar field \citep{MP03,Lilley08}. Still in
the classical domain, one can either modify gravity in order to render
it non singular \citep{Fabris,APY10} or consider a fluid that does not
satisfy the usual energy conditions \citep{AP07,FPP08}. Finally, one
can use some version of quantum gravity that would apply to the
Universe as a whole: these would lead to quantum cosmology models
\citep{PPP05,PPP06,PPP07}, allowing for a bounce independently of the
curvature. In almost all of these models, it turns out that the
spectrum of primordial perturbations is modified to include an
oscillatory part, that can be compared to observations as examplified
on Fig. \ref{FigFLP}.

\section{Conclusions}

Precision cosmology is opening new windows of observations: future
data will constrain high energy theories!

\begin{acknowledgements}
I should like to thank J.~Martin and C.~Ringeval for many enlightning
discussions and for providing the relevant figures coming from their
works.
\end{acknowledgements}


\begin{thebibliography}{}
\bibitem[Abramo \& Peter (2007)]{AP07}Abramo, L., R.\& Peter, P.,
  2007,  JCAP 09, 001.
\bibitem[Abramo et al.(2010)]{APY10} Abramo, L., R., Peter,
  P.  \& Yasuda, I., 2010, Phys. Rev. D81, 023511.
\bibitem[Abazajian et al.(2009)]{LSS} Abazajian, K., et al. 2009, Ap. J. S., 182, 543.
\bibitem[Amanullah et al.(2009)]{SNIa} Amanullah, R. et al., 2010, Ap. J. 716, 712.
\bibitem[Bouchet et al.(2002)]{BPRS} Bouchet, F. R., Peter,  P.,
  Riazuelo, A. \& Sakellariadou, M., 2001, Phys. Rev. D65, 021301.
\bibitem[Fabris et al.(2003)]{Fabris} Fabris, J.~C., Furtado, R.~G.,
  Peter, P. \& Pinto-Neto, N., 2003, Phys. Rev. D67, 124003.
\bibitem[Falciano et al.(2008)]{Lilley08} Falciano, F. T., Lilley, M.
  \& Peter, P. Phys. Rev. D77, 083513.
\bibitem[Finelli et al.(2008)]{FPP08} Finelli, F.,
  Peter, P. \& Pinto-Neto, N., 2008, Phys. Rev. D77, 103508.
\bibitem[Fraisse et al.(2008)]{FRSB} Fraisse, A. A., Ringeval, C.,
  Spergel D. N.  \& Bouchet, F. R., 2008, Phys. Rev. D78, 043535.
\bibitem[Furlanetto et al.(2006)]{21cm} Furlanetto, S.~R., Peng Oh, S. \& Briggs,
  F.~H., 2006, Phys. Rep. 433, 181.
\bibitem[Jeannerot et al.(2003)]{rocher} Jeannerot, R., Rocher, J. \&
  Sakellariadou, M., 2003, Phys. Rev. D68, 103514
\bibitem[Komatsu(2010)]{WMAP7} Komatsu, E. et al., 2010, arXiv:1001.4538.
\bibitem[Martin(2008)]{JM08} Martin, J., 2008, Lect. Notes Phys. 738, 193.
\bibitem[Martin \& Peter(2003)]{MP03} Martin, J., \& Peter, P., 2003,
  Phys. Rev. D68, 103517.
\bibitem[Martin \& Peter(2004)]{MP04} Martin, J., \& Peter, P., 2004,
  Phys. Rev. Lett. 92, 061301.
\bibitem[Martin \& Ringeval(2010)]{JMCR10} Martin, J. \& Ringeval, C.,
  2001, Phys. Rev. D82, 023511.
\bibitem[Martin et al.(2010)]{MRT10} Martin, J.,
  Ringeval, C., Trotta, R., 2010 arXiv:1009.4157
\bibitem[Percival et al.(2009)]{BAO} Percival, W.~J. et al., 2010,
  MNRAS 401, 2148.
\bibitem[Peter \& Pinto-Neto(2002)]{PPNPN02} Peter, P. \& Pinto-Neto,
  N., 2002, Phys. Rev. D66, 063509.
\bibitem[Peter \& Pinto-Neto(2008)]{PPNPN} Peter, P. \& Pinto-Neto,
  N., 2008, Phys. Rev. D78, 063506.
\bibitem[Peter et al.(2005)]{PPP05} Peter, P. Pinho, E.,
  \& Pinto-Neto, 2005, JCAP 07, 014.
\bibitem[Peter et al.(2006)]{PPP06} Peter, P. Pinho, E.,
  \& Pinto-Neto, 2006, Phys. Rev. D73, 104017.
\bibitem[Peter et al.(2007)]{PPP07} Peter, P. Pinho, E.,
  \& Pinto-Neto, 2007, Phys. Rev. D75, 023516 (2007).
\bibitem[Peter and Uzan(2009)]{PU09} Peter, P. \& Uzan, J.-P., 2009,
  Primordial Cosmology (Oxford University Press).
\bibitem[Pogosian et al.(2009)]{Pog} Pogosian, L., Henry Tye, S.-H.,
  Wasserman, I. \& Wyman, M., 2009, JCAP 0902, 013.
\bibitem[Ringeval(2010)]{CR10} Ringeval, C., 2010, arXiv:1005.4842
\bibitem[Vilenkin \& Shellard(2000)]{VilShe00} Vilenkin, A. \&
  Shellard, E.~P.~S., 2000, Cosmic Strings and Other Topological Defects (CUP).
\end{thebibliography}
\end{document}